\documentclass[aps,a4paper,10pt]{revtex4}

\usepackage{amsmath}
\usepackage{amssymb}
\usepackage{lineno}
\usepackage{graphicx}
\usepackage{subfigure}
\def\tens{\overleftrightarrow}

\begin{document}



\title{
       {\parbox[b]{\textwidth}{\rm \footnotesize \begin{flushright}
       K. H. Hughes (ed.) \\
       Dynamics of Open Quantum Systems \\
       \copyright \hspace{1ex} 2010, CCP6, Daresbury
        \end{flushright}}} \\
Quantum dynamics and super-symmetric quantum mechanics.}


\author{Eric R. Bittner}
\email[corresponding author]{}
\affiliation{Department of Chemistry, University of Houston, Houston, TX  77204}

\author{Donald J. Kouri}
\affiliation{Department of Chemistry, University of Houston, Houston, TX  77204}


\date{\today}

\begin{abstract}
In my talk I will present an overview of our recent work involving the use of 
supersymmetric quantum mechanics (SUSY-QM).
I begin by discussing the mathematical underpinnings of SUSY-QM and then 
discuss how we have used this for developing novel theoretical and numerical 
approaches suitable for studying molecular systems.  I will conclude by discussing our 
attempt to extend SUSY-QM to multiple dimensions.
\end{abstract}

\pacs{}

\maketitle
\setcounter{page}{1}
\thispagestyle{plain}


\section{A first date}

My introduction to supersymmetric quantum mechanics (SUSY) came quite by accident.  
 I  had heard of SUSY in the context of high-energy physics where the SUSY theory 
 postulates that for every fermion there is  boson of equal mass (i.e. energy).  This comes about
because for every quantum Hamiltonian there is a partner Hamiltonian that has the 
same energy spectrum above the ground-state of the original system.  In other words, 
above the ground state of $H_{1}$, each higher-lying eigenstate is partnered with 
an eigenstate of $H_{2}$.  In particle physics, the $H_{1}$ ``sector''  is populated by bosons
and the $H_{2}$ sector by fermions and SUSY predicts that the lowest lying fermion state
is energetically degenerate with the first excited boson state.   Evidence for SUSY has proven to 
be elusive and it is  now believed that SUSY is a broken symmetry.  

Last January (Jan-09) at a conference dedicated to  Bob Wyatt, my co-author suggested that 
we look at SUSY as a way to develop new computational methods and approaches.  
Up until now, SUSY has been more of a mathematical technique that has been used more or less 
as a way to obtain stationary solutions to the Schr\"odinger equation for
 the variety of one-dimensional potential systems.   In this paper and in my talk, I will discuss some of the work we have been doing in developing 
 ``SUSY'' inspired methods for performing quantum many-body calculations and quantum scattering calculations. 
 I shall begin with a brief overview of the SUSY theory and some of its elementary results. I shall then 
 discuss how we have used the approach to develop both analytical and numerical solutions 
 of the stationary Schr\"odinger equation.  I will conclude by discussing our recent extension of SUSY to 
 higher dimensions and for scattering theory. 

\section{Mathematical considerations} 

Before discussing some of our recent results, it is important to introduce briefly  the 
mathematical formulation of SUSY quantum mechanics.  

\subsection{Hamiltonian formulation of SUSY}
In quantum theory, there is a fundamental connection between a bound state and its potential. 
This is simple to demonstrate by writing the Schr\"odinger equation
for the stationary states  as
\begin{eqnarray}
V_{1}(x)  - E_{n}  = -\frac{\hbar^2}{2m}\frac{1}{\psi_{n}}\partial_x^{2}\psi_{n} = Q[\psi_{n}]
\end{eqnarray}
where we recognize the right-hand side as the Bohm quantum potential which will certainly be
discussed repeatedly at this conference. 
 One of the remarkable consequences of this equation is that
every stationary state of a given potential has the same functional form for its
quantum potential $Q$. Thus, knowing any bound state allows 
a global reconstruction of the potential, $V(x)$ up to a constant energy shift. 

SUSY is obtained 
by factoring
the Schr\"odinger equation into the form \cite{Witten:1981bq,Witten:1982nx,Cooper:1995jt}
\begin{eqnarray}
H\psi = A^{+}A \psi_{o}^{(1)} = 0 \label{susy-hamiltonian}
\end{eqnarray}
using the operators
\begin{eqnarray}
A  = \frac{\hbar}{\sqrt{2m}} \partial_{x}+ W  \,\,{\rm  and}\,\,
A^{+}  = -\frac{\hbar}{\sqrt{2m}}\partial_{x} + W .
\end{eqnarray}
Since we can impose  $A\psi_{o}^{(1)} = 0$, we can immediately write that 
\begin{eqnarray}
W(x) = -\frac{\hbar}{\sqrt{2m}}\partial_x \ln\psi_{o}.
\end{eqnarray}
$W(x)$ is the {\em superpotential} which is related to the physical 
potential by a Riccati equation.
\begin{eqnarray}
V(x) = W^{2}(x)  - \frac{\hbar}{\sqrt{2m}}W'(x). \label{re}
\end{eqnarray}
The SUSY factorization of the Schr\"odinger equation can always be applied in one-dimension.

From this point on we label the 
original Hamiltonian operator and its associated potential, states, and energies as $H_{1}$, $V_{1}$, 
$\psi_{n}^{(1)}$  and $E_{n}^{(1)}$.
One can also
define a partner Hamiltonian, $H_{2} = AA^{+}$ with a corresponding potential
\begin{eqnarray}
V_{2} = W^{2} +  \frac{\hbar}{\sqrt{2m}}W'(x).
\end{eqnarray}
All of this seems rather circular and pointless until one recognizes that $V_{1}$ and its partner potential, 
 $V_{2}$,  give rise to a common set of
energy eigenvalues. 
This principle result of SUSY 
can be seen by first considering an arbitrary stationary solution of $H_{1}$,
\begin{eqnarray}
H_{1} \psi_{n}^{(1)} = A^{+}A\psi_{n} = E_{n}^{(1)}\psi_{n}^{(1)}.
\end{eqnarray}
This implies that $(A\psi_{n}^{(1)})$ is an eigenstate of $H_{2}$ with energy $E_{n}^{(1)}$ since
\begin{eqnarray}
H_{2}(A\psi_{n}^{(1)}) = AA^{+}A\psi_{n}^{(1)} = E_{n}^{(1)}(A\psi_{n}^{(1)}).
\end{eqnarray}
Likewise, the Schr\"odinger equation involving the partner potential $H_{2}\psi_{n}^{(2)} = E_{n}^{(2)}\psi_{n}^{(2)} $ implies that
\begin{eqnarray}
A^{+}AA^{+}\psi_{n}^{(2)} = H_{1}(A^{+}\psi_{n}^{(2)}) = E_{n}^{(2)}(A^{+}\psi_{n}^{(2)}).\label{chargeop}
\end{eqnarray}
This (along with $E_{o}^{(1)} = 0$ ) allows one to conclude that the eigenenergies and eigenfunctions of $H_{1}$ and 
$H_{2}$ are related in the following way: 
$
E_{n+1}^{(1)} = E_{n}^{(2)},
$
\begin{eqnarray}
\psi_{n}^{(2)} = \frac{1}{\sqrt{E_{n+1}^{(1)}}} A \psi_{n+1}^{(1)},\, \,{\rm and} \,\,
\psi_{n+1}^{(1)} = \frac{1}{\sqrt{E_{n}^{(2)}}} A^{+} \psi_{n}^{(2)}  \label{ops}
\end{eqnarray}
for $n > 0$.  
\footnote{Our notation from here on is that $\psi_{n}^{(m)}$ denotes the $n$th state associated with the $m$th partner Hamiltonian 
with similar notion for related quantities such as energies and superpotentials. }
Thus, the {\em ground state of $H_{2}$ has the same energy as the first excited state of $H_{1}$}.
If this state $\psi_{o}^{(2)}$ is assumed to be node-less, then  $\psi_{1}^{(1)}  \propto A^{+}\psi_{o}^{(2)}$
will have a single node.     We can repeat this analysis and show that $H_{2}$ is partnered with another 
Hamiltonian, $H_{3}$ whose ground state is isoenergetic with the first excited state of $H_{2}$ and thus
isoenergetic with the second excited state of the original $H_{1}$.  This hierarchy of  partners
persists until all of the bound states of $H_{1}$ are exhausted.

\subsection{SUSY algebra}
We can connect the two partner Hamiltonians by constructing a matrix super-Hamiltonian operator
\begin{eqnarray}
{\bf H} =
\left(
\begin{array}{cc}
H_1  & 0 \\
0  & H_2 \\
\end{array}
\right)
\end{eqnarray}
and two matrix ``super-charge'' operators
\begin{eqnarray}
{\bf Q} 
=
\left(
\begin{array}{cc}
0  & 0 \\
A  & 0\\
\end{array}
\right) 
= A \sigma_{-} \end{eqnarray}
and
\begin{eqnarray}
{\bf Q}^{+} 
=
\left(
\begin{array}{cc}
0  & A^{+} \\
0  & 0\\
\end{array}
\right) = A^{+} \sigma_{+}
\end{eqnarray}
where $\sigma_{\pm}$ are $2\times 2$ Pauli spin matrices.  
Using these we can re-write the SUSY Hamiltonian as 
\begin{eqnarray}
{\bf H} 
= 
\left(-\frac{\hbar^{2}}{2m}\frac{d^{2}}{dx^{2}} + W^{2}\right) \sigma_{o}
+W'\sigma_{z}
\end{eqnarray}
The operators $\{ {\bf H}, {\bf Q}, {\bf Q}^{+}\} $ form a closed algebra (termed the Witten superalgebra)
with
\begin{eqnarray}
[ {\bf H}, {\bf Q}  ] &=& [ {\bf H}, {\bf Q}^{+} ] = 0  \\
\{ {\bf Q},  {\bf Q} \} &=& \{ {\bf Q}^{+},  {\bf Q}^{+} \} = 0 \\
\{ {\bf Q},  {\bf Q}^{+} \} &=& {\bf H}
\end{eqnarray}
The first algebraic relation is responsible for the degeneracy of the spectra of $H_{1}$ and $H_{2}$ and the 
supercharges transform an eigenstate of one sector into an eigenstate of the other sector.

As an example and perhaps a better connection to the physics implied by this structure, consider the case 
of a one-dimensional particle with an internal spin degree of freedom and 
with $[x,p] = i $ denoting the position and momentum of the particle. 
Conserved SUSY would imply that all non-diagonal coupling terms between the bosonic (coordinate) and 
fermionic (spin) degrees of freedom are exactly zero. 
This  of course is equivalent to making the 
 Born Oppenheimer approximation for a two-state system coupled to a continuous field $x(t)$.   
 In this case,  SUSY is preserved so long as $d_{t}\psi(x(t),t) = \partial_{t} \psi(x(t),t)$.  
 SUSY is broken when $\dot x(t) \partial_{x}\psi(x(t),t) \ne 0$ which would lift the degeneracy 
 between the states of $H_{1}$ and $H_{2}$.

\subsection{Scattering in one dimension}

The SUSY approach is not limited to bound-state problems. 
For a one-dimensional scattering system, it is straightforward to apply the SUSY theory to determine a relation between 
between the transmission and reflection coefficients of the supersymmetric partners.  Asymptotically, 
 we can assume that $W(x) \to W_{\pm}$ as $x\to \pm \infty$. In the same limit, the partner potentials
become $V_{1,2} \to W_{\pm}^{2}$.   For a plane wave incident from the left with energy $E$ scattering from $V_{1,2}$, we
require the following asymptotic forms:
\begin{eqnarray}
\lim_{x\to-\infty}\psi^{(1,2)}(k,x) &\sim &e^{ikx} +  R^{(1,2)}e^{-ikx}  \\
\lim_{x\to+\infty}\psi^{(1,2)}(k',x)&\sim& T^{(1,2)}e^{ik'x}  
\end{eqnarray}
We can derive a relation between the two scattering states by using the relation $\psi^{(1)}(k,x) = N A^{+}\psi_{2}(k',x)$.
For the left-hand components  ($x\to -\infty$).
\begin{eqnarray}
e^{ikx}+R^{(1)}e^{-ikx} = N\left[
\left(-{i k } + \tilde W_{-}\right)e^{ikx}
+\left({i k } + \tilde W_{-}\right)e^{-ikx}
\right]
\end{eqnarray}
where in the last line we have incorporated the $\hbar/\sqrt{2m}$ in to the normalization and wrote
 $\tilde W_{\pm} =W_{\pm} \sqrt{2m}/{\hbar}$.
Likewise for the transmitted coefficients ($x\to +\infty$).
\begin{eqnarray}
T^{(1)}e^{ik'x} = N (-ik' + \tilde W_{+} )T^{(2)}e^{ik'x}
\end{eqnarray}
Eliminating the common normalization factor and using the fact that $k = \sqrt{2m(E-W_{-})}/\hbar$ and 
 $k' = \sqrt{2m(E-W_{+})}/\hbar$  from the Schr\"odinger equation we can arrive at 
\begin{eqnarray}
R^{(1)}(k) = \frac{W_{-}+ik}{W_{-}-ik}R^{(2)}(k) \\
T^{(1)}(k) = \frac{W_{+}-ik'}{W_{-}-ik}T^{(2)}(k).
\end{eqnarray}
Consequently, knowledge of the scattering states of $V_{1}$ allows one to easily construct 
scattering states for the partner potential. 

\subsection{Non-stationary states}
Finally, one can use the SUSY approach in a time-dependent context by writing
$$
i\hbar \partial_{t} \psi^{(1)} = H_{1} \psi^{(1)} = A^{+} A \psi^{(1)}
$$
where $\psi^{(1)}$ is a non-stationary state in the first sector.  If $V_{1}$ is independent of time, 
then the superpotential must also be independent of time and so we can write
$$
i\hbar A \partial_{t} \psi^{(1)} =  i\hbar \partial_{t} (A \psi^{(1)}) =  A A^{+} (A  \psi^{(1)}) 
$$
In other words, we have the time-dependent Schr\"odinger equation for the partner potential
$$i\hbar \partial_{t}\psi^{(2)} =   H_{2} \psi^{(2)}.
$$
The two non-stationary states are partnered, $\psi^{(2)} \propto A^{+}\psi^{(1)}$. 
We also note that these states satisfy
$$
\psi^{(1)}(t) = e^{-iA^{+}At/\hbar}\psi(0)
$$
and
$$
\psi^{(2)}(t) = e^{-iAA^{+}t/\hbar}\psi(0)
$$
for some initial state $\psi(0)$.  
Using the charge operators we can show that
$$
A\psi^{(1)}(t) = 
e^{-iAA^{+}t/\hbar}(A\psi(0)).
$$
As above in the scattering example, one can use the dynamics of one sector to determine the dynamics in the other sector.

The partnering scheme presents a powerful prescription for developing novel approaches for solving a wide variety of 
quantum mechanical problems. 
This allows one one use analytical or numerical solutions of one problem to determine either approximate or exact solutions to some new problem. 
 In the sections that follow, I present some of our attempt to use SUSY in a 
numerical context.   At the moment our numerical results are limited to one spatial dimension.  As I shall 
discuss, extending SUSY to multiple dimensions has proven to be problematic. However, in Sec. V we 
present our extension using a vector-SUSY approach we are developing.

\section{Using SUSY to obtain excitation energies and excited states}

The SUSY hierarchy also provides a useful prescription for determining the excited states of $H_{1}$ (which may represent the physical problem of interest.) The first excited state of $H_{1}  $ is isoenergetic with the ground state of $H_{2}$. 
Since this state is node-less, one can use either Ritz variational approaches or Monte Carlo approaches to determine this 
state to very high accuracy. 

Two basic tools used in computational chemistry are the Quantum Monte
Carlo (QMC) and the Rayleigh-Ritz variational approaches. Both
approaches yield their best and most accurate results for ground state
energies and wave functions. Although the variational method also
gives bounds for the excited state energies as well as the ground
state (the Hylleraas-Undheim theorem \cite{Hilleraas:1930ph}), it is
well known that their accuracy is significantly lower than that of the
ground state. Even more serious, the wave functions are known to
converge much more slowly than the energies.

In the case of the
QMC\cite{Hammond94,porter:7795,Doll87,Needs:2001kx,PhysRevB.71.241103,PhysRevE.55.3664},
there are additional difficulties associated with the presence of
nodes in the excited state wave functions \cite{bouabca:114107}. While
some progress has been made in dealing with this issue (e.g., the
``fixed node'' or ``guide wave''
techniques)\cite{Needs:2001kx,PhysRevB.71.241103,PhysRevE.55.3664,bouabca:114107,Oriols98}
the computational effort required is greater and the accuracy is lower
and in fact, no general solution to the difficulty has been found for
reducing the computational effort and increasing the accuracy for
excited state calculations in QMC to the same level as is attained for
the ground state.  In fact, it is very likely the presence and effects
of nodes in the excited states that is largely responsible for the
lower accuracy and slower convergence of excited state results in the
variational method. The precise determination of nodal surfaces is
expected to play a crucial role since they reflect changes in the
relative phase of the wave function. Because of the ubiquitous
importance of both the variational and QMC methods, solving the
so-called ``node problemÓ will have enormous impact on computational
chemistry.

\subsection{Using SUSY to improve quality of variational calculations}

We now turn to the proof of principle for this approach as a computational scheme to obtain improved excited state energies and wave functions in the Rayleigh-Ritz variational method. We should note that these results can be generalized to any system where a hierarchy of Hamiltonians can be generated because of the nature of the Rayleigh-Ritz scheme. In the standard approach one calculates the energies and wave functions variationally, relying on the Hylleraas-Undheim theorem for convergence\cite{Hilleraas:1930ph}. This, however, is unattractive for higher energy states because they require a much larger basis to converge to the same error. We stress that this is true regardless of the specific basis set used. Of course, some bases will be more efficient than others but it is generally true that for a given basis, the Rayleigh-Ritz result is less accurate for excited states. We address this situation by solving for ground states in the variational part of the problem.

\begin{figure}[t]
\includegraphics[width=0.5\columnwidth]{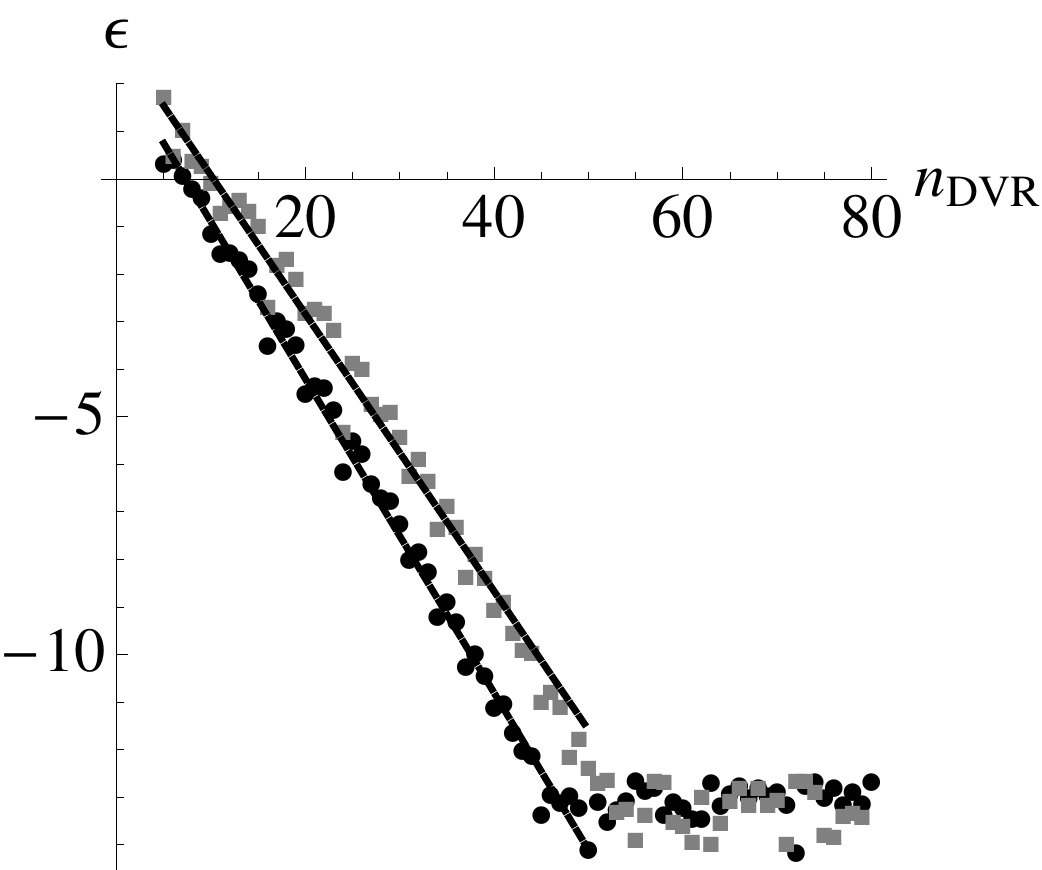}
\caption{Convergence of first excitation energy $E_{1}^{1}$ for model potential $V_{1} = x^{6} + 4 x^{4} + x^{2} -2$ 
using a $n$-point discrete variable representation (DVR).  Gray squares: $\epsilon=\log_{10}|E_{1}^{1}(n)-E_{1}^{1}(exact)|$,
Black squares:   $\epsilon=\log_{10}|E_{0}^{2}(n)-E_{1}^{1}(exact)|$.  Dashed lines are linear fits. 
(From Ref\cite{Kouri:2009hb}.)}\label{converge}
\end{figure}

To demonstrate our computational scheme, we investigate the first example system from the previous section. For the potential 
\begin{equation}
V_1(x) = x^6 + 4x^4 + x^2 - 2.
\end{equation}
exact solutions are known for all states of $H_1$. We use the exact results to assess the accuracy of the variational calculations.   Here we employed a $n$-point discrete variable representation (DVR)  based upon the 
Tchebchev polynomials to compute the eigenspectra  of the first  and second sectors.\cite{light:1400,lig92:185}  In Fig. \ref{converge} we show the numerical error in the first excitation energy by comparing $E_{1}^{1}(n)$ and $E_{0}^{2}(n)$ 
from an $n$ point DVR to the numerically ``exact''  value corresponding to a 100 point DVR,
$$\epsilon_{1}^{1}(n) = \log_{10}|E_{1}^{1}(n)-E_{1}^{1}(exact)|.$$ 
Likewise,
$$\epsilon_{0}^{2}(n) = \log_{10}|E_{0}^{1}(n)-E_{1}^{1}(exact)|.$$
For any given basis size, $\epsilon_{0}^{2} < \epsilon_{1}^{1}$.  Moreover, over a range of $15 < n < 40$ points, the excitation energy computed using the 
second sector's ground state is between 10 and 100 times more accurate than $E_{1}^{1}(n)$.
This effectively reiterates our point that by using the SUSY hierarchy, one can systematically improve upon the accuracy
of a given variational calculation.

\subsection{Monte Carlo SUSY }
Having defined the basic terms of SUSY quantum mechanics, let us presume that one can determine  
an accurate approximation to the ground state density $\rho_{o}^{(1)}(x)$  of Hamiltonian $H_{1}$.
One can then  
use this to determine the superpotential using the Riccati transform 
\begin{eqnarray}
W_{o}^{(1)} = -\frac{1}{2} \frac{\hbar}{\sqrt{2m}} \frac{\partial \ln\rho_{o}^{(1)}}{\partial x} \label{riccati}
\end{eqnarray}
and the partner potential
\begin{eqnarray}
V_{2} = V_{1} - \frac{\hbar^{2}}{2m} \frac{\partial^{2} \ln\rho_{o}^{(1)}}{\partial x^{2}}. \label{eq13}
\end{eqnarray}
Certainly, our ability to compute the energy of the ground state of the partner 
potential $V_{2}$ depends on having first obtained an accurate estimate of the 
ground-state density associated with the original $V_{1}$.   

For this we turn to an adaptive Monte Carlo-like approach developed by 
Maddox and Bittner.\cite{maddox:6465}  Here, we assume we
can write the trial density as a 
sum over $N$  Gaussian approximate functions
\begin{eqnarray}
\rho_{T}(x) = \sum_{n} G_{n}(x,{\bf c}_{n}). \label{approx}
\end{eqnarray}
parameterized by their amplitude, center, and width.    
\begin{eqnarray}
G_{n}(x,\{{\bf c}_{n}\}) = c_{no} e^{-c_{n2}(x-c_{n3})^{2}}
\end{eqnarray}
This trial density then is used to compute the energy
\begin{eqnarray}
E[\rho_{T}] = \langle V_{1}\rangle + \langle Q[\rho_{T}]\rangle 
\end{eqnarray}
where $ Q[\rho_{T}] $ is the Bohm quantum potential,
\begin{eqnarray}
Q[\rho_{T} ] = -\frac{\hbar^{2}}{2m}\frac{1}{\sqrt{\rho_{T}}}\frac{\partial^{2}}{\partial x^{2}}\sqrt{\rho_{T}}.
\end{eqnarray}
  The energy average 
is computed by sampling $\rho_{T}(x)$ over a set of trial points $\{x_{i}\}$ and then 
moving the trial points along the conjugate gradient of 
\begin{eqnarray}
E(x) = V_{1}(x) + Q[\rho_{T}](x).
\end{eqnarray}
After each conjugate gradient step, a new set of $\bf {c}_{n}$ coefficients are determined 
according to an expectation maximization criteria such that the new trial density provides
the best $N$-Gaussian approximation to the actual probability distribution function sampled by the
new set of trial points.  The procedure is repeated until $\delta \langle E\rangle = 0$. In doing so, we simultaneously 
minimize the energy and optimize the trial function.
Since the ground state is assumed to be node-less, we will not encounter the singularities and 
numerical instabilities associated with other Bohmian equations of motion based approaches. 
 \cite{Bohm52a,Holland93,Lopreore99,Bittner00a,Wyatt01,maddox:6465}
Moreover, the approach has been extended to very high-dimensions and to finite temperature
by Derrickson and Bittner in their studies of the structure and thermodynamics of 
rare gas clusters with up to 130 atoms.  \cite{Derrickson:2006,Derrickson:2007jo}

\section{Test case: tunneling in a double well potential}
As a non-trivial test case, consider the tunneling of a particle 
between two minima of a symmetric double potential well.
One can estimate the tunneling splitting using semi-classical techniques by assuming that the ground and 
excited states are given by the approximate form
\begin{eqnarray}
\psi_{\pm} = \frac{1}{\sqrt{2}}(\phi_{o}(x) \pm \phi_{o}(-x))
\end{eqnarray}
where $\phi_{o}$ is the lowest energy state in the right-hand well in the limit the wells are 
infinitely far apart.  
From this, one can easily estimate the splitting as \cite{Landau:1974wq}
\begin{eqnarray}
\delta = 4 \frac{\hbar^{2}}{m} \phi_{o}(0)\phi_{o}'(0)
\end{eqnarray}
If we assume the localized states $(\phi_{o})$ to be gaussian,  then 
\begin{eqnarray}
\psi_{\pm} \propto \frac{1}{\sqrt{2}}(e^{-\beta(x-x_{o})^{2}}\pm e^{-\beta(x+x_{o})^{2}})
\end{eqnarray}
and we can write the superpotential as 
\begin{eqnarray}
W = \sqrt{\frac{2}{m}}\hbar\beta \left(x - x_{o}\tanh(2 x x_{o}\beta) \right).
\end{eqnarray}
From this, one can easily determine both the original potential and the partner potential as
\begin{eqnarray}
V_{1,2} &=& W^{2} \pm \frac{\hbar}{\sqrt{2m}}W' \\
&=& \frac{\beta^{2}  \hbar ^2}{m} \left( 
2   (x-x_o \tanh (2 x x_o \beta ))^2 \right. \nonumber \\
&\pm&\left. (2 x_o^2   \text{sech}^2(2 x x_o \beta )-1\right)
\end{eqnarray}
While the $V_{1}$ potential has the characteristic double minima giving rise to a tunneling doublet, 
the SUSY partner potential $V_{2}$ has a central dimple which in the limit of $x_{o}\rightarrow \infty$ becomes
a $\delta$-function which produces an unpaired and node-less ground state. \cite{Cooper:1995jt}
Using Eq.~\ref{chargeop}, one obtains $\psi_{1}^{(1)} = \psi_{-} \propto A^{\dagger}\psi_{o}^{(2)}$ which now has a single node
at $x = 0$.

\begin{figure}[t]
\subfigure[]{\includegraphics[width=0.25\columnwidth]{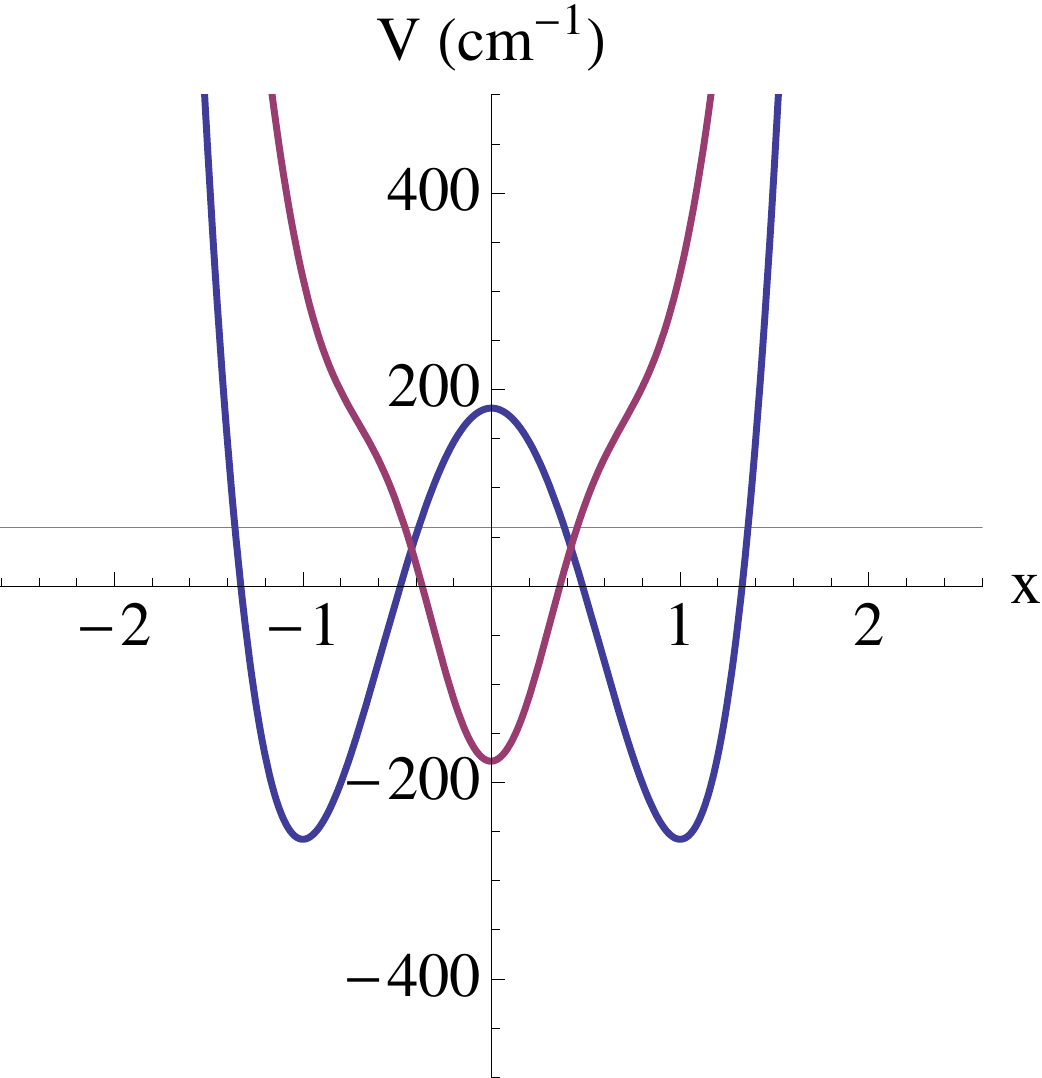}}
\subfigure[]{\includegraphics[width=0.25\columnwidth]{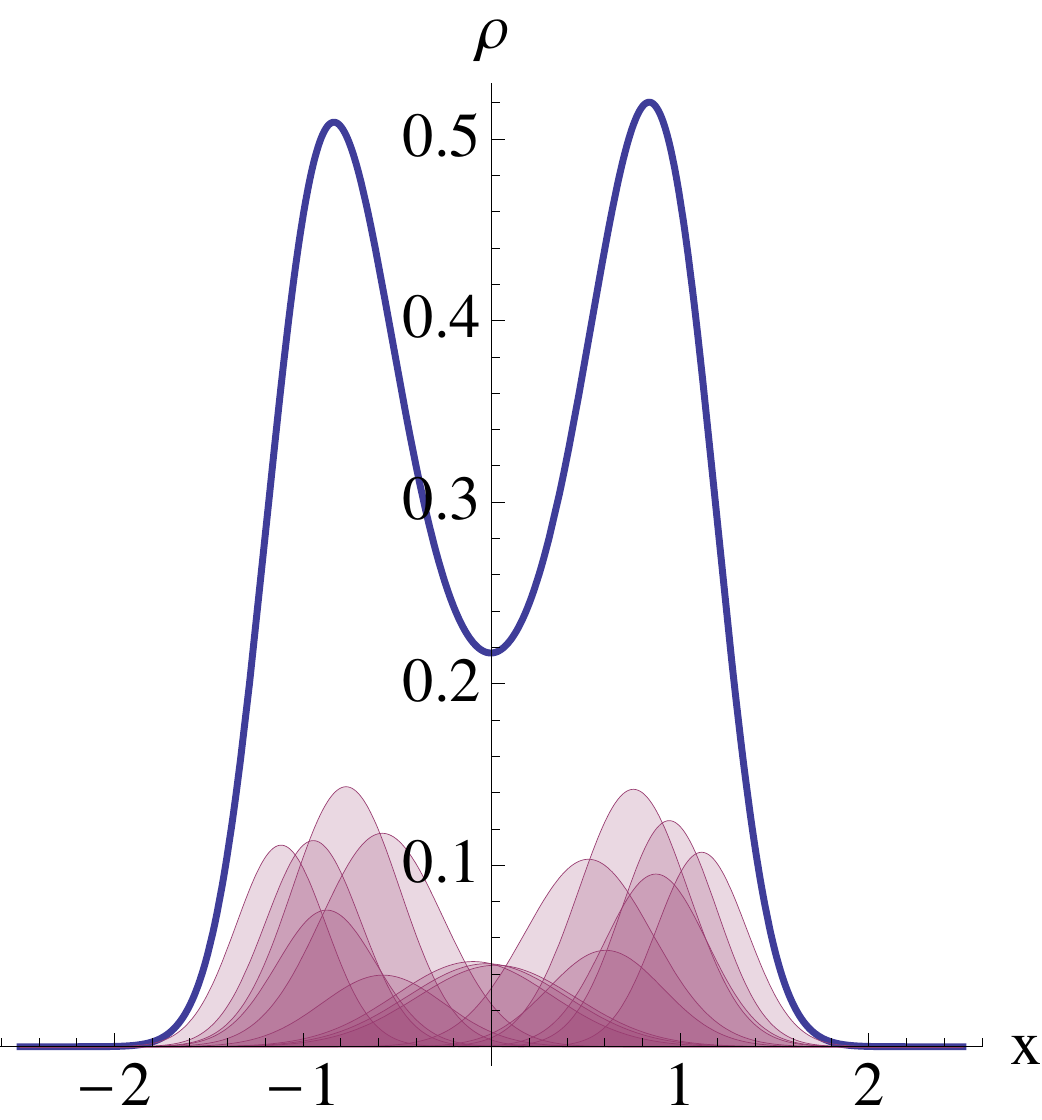}}
\caption{(a) Model double well potential(blue)  and partner potential (purple). The energies of the 
tunneling doublets are indicated by the horizontal lines at $ V = 0\, {\rm  cm}^{-1}$ and $V =59.32\, {\rm  cm}^{-1}$ 
indicating the positions of the sub-barrier tunneling doublet. 
(b) Final ground state density (blue) superimposed over the Gaussians used in its expansion. (purple)
(From Ref.\cite{Bittner:2009jt})
}\label{fig1}
\end{figure}

For a computational example,  we take the double well potential to be of the form
\begin{eqnarray}
V_{1}(x) = a x^{4} + bx^{2} + E_{o}.
\end{eqnarray}
with  $a = 438.9 {\rm cm}^{-1}/(bohr^{2})$, $b = 877.8 {\rm cm}^{-1}/(bohr)^{4}$, and
$E_{o} = -181.1 {\rm cm}^{-1}$
which (for $m = m_{H}$ ) 
gives rise to exactly two states at  below the barrier separating 
the two minima with a tunneling splitting of 
59.32 ${\rm  cm}^{-1}$ 
as computed using a discrete variable representation (DVR) approach.\cite{lig85:1400}
 For the calculations reported here, 
we used $n_{p}=1000$ sample points and $N =15$ Gaussians and  in the expansion of $\rho_{T}(x)$ 
to converge the ground state.   
This converged the ground state to $1:10^{-8}$ in terms of the energy.
This is  certainly a bit of an overkill in the number of points and number of gaussians since far 
fewer DVR points were required to achieve comparable accuracy (and a manifold of excited states). 
The numerical results, however, are encouraging since the accuracy of  generic Monte Carlo 
evaluation would be $1/\sqrt{n_{p}} \approx 3\%$ in terms of the energy.  
\footnote{In our implementation, the sampling points are only used to 
evaluate the requisite integrals and they themselves are 
adjusted along a conjugate gradient rather than by resampling. 
One could in principle forego this step entirely and optimize the 
parameters describing the gaussians directly.   }
Plots of $V_{1}$ and the converged ground state is shown in ~\ref{fig1}.   
 
The partner potential  $V_{2} = W^{2} + \hbar W'/\sqrt{2m}$, 
can be constructed once we know the superpotential, $W(x)$.   
Here, we require an accurate evaluation of the ground state density and 
its first  two log-derivatives.   The advantage of our computational scheme is that one 
can evaluate these analytically for a given set of coefficients.  
In  \ref{fig1}a we show the partner potential derived from the ground-state density. 
Where as the original  $V_{1}$
 potential exhibits the double well structure with minima near $x_{o} = \pm 1$ ,
  the $V_{2}$ partner potential has a pronounced dip about $x=0$.    Consequently, 
  its ground-state should have a simple ``gaussian''-like form peaked about the origin. 

\begin{figure}[t]
\subfigure[]{\includegraphics[width=0.25\columnwidth]{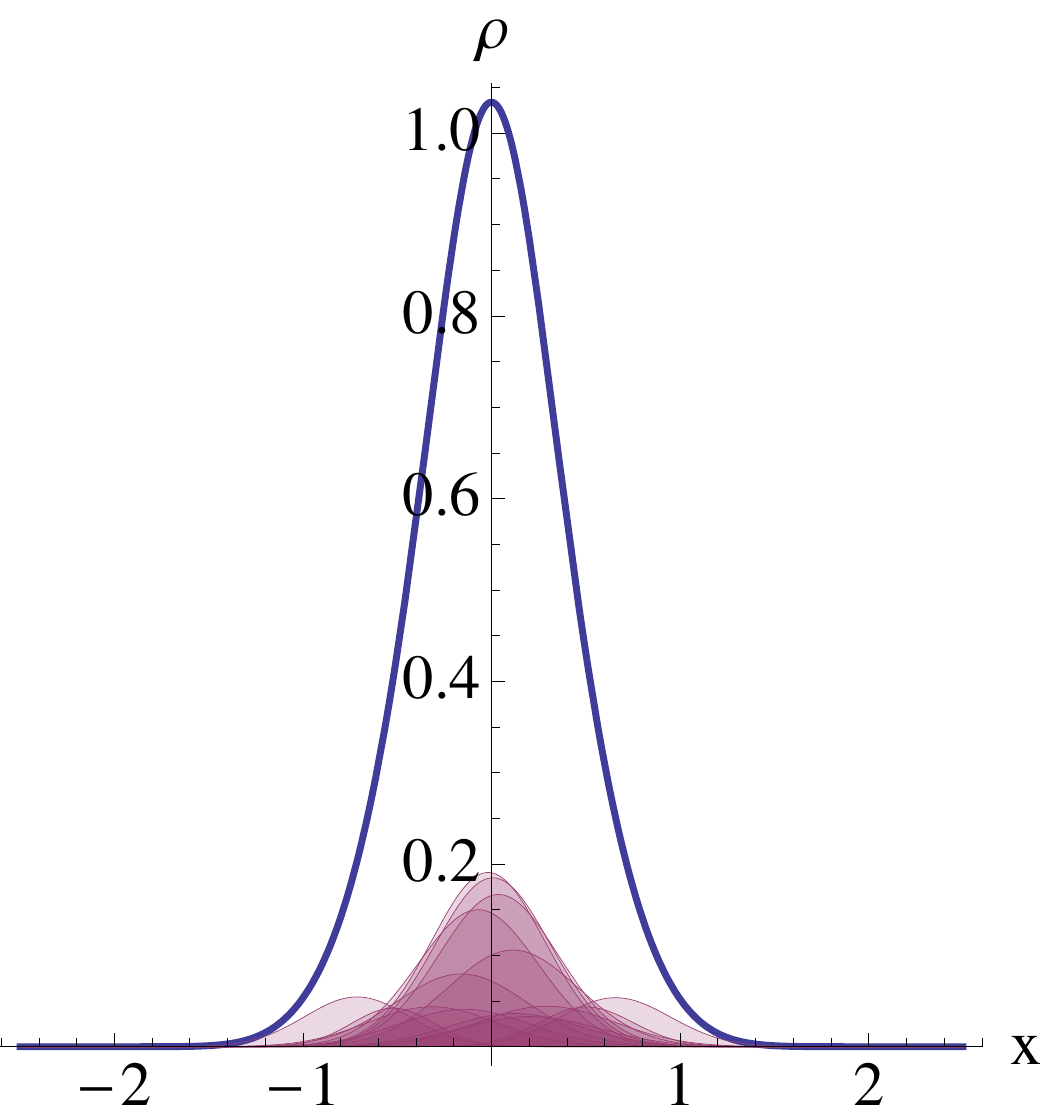}}
\subfigure[]{\includegraphics[width=0.25\columnwidth]{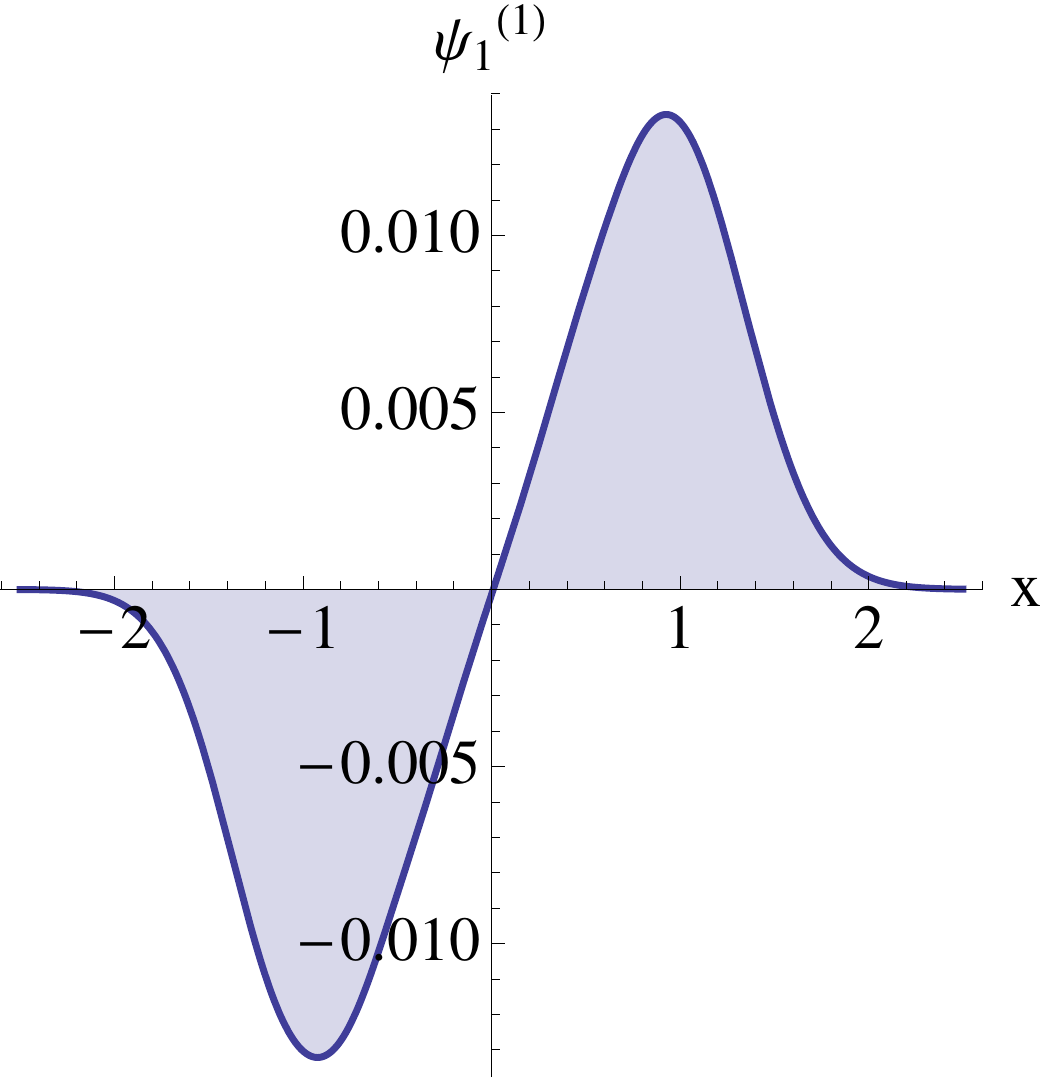}}
\caption{(a) Ground state density of the partner Hamiltonian $H_{2}$ (blue) superimposed over its
individual Gaussian components.  (b) 
Excited state  $\psi_{1}^{(1)}$ derived from the ground state of the partner potential, $\psi_{o}^{(2)}$.
(From Ref.\cite{Bittner:2009jt})
 }\label{fig4}
\end{figure}

Once we determined an accurate representation of the partner potential, it is now a trivial matter to
re-introduce the partner potential into the optimization routines.
The ground state converges easily and is shown in ~\ref{fig4}a along with its
gaussians. After 1000 CG steps, the converged energy is within 0.1\% of the exact tunneling splitting for this model system.
Again, this is an order of magnitude better than the $1/\sqrt{n_{p}}$ error associated with a simple Monte Carlo sampling. 
Furthermore, ~\ref{fig4}b shows  $\psi_{1}^{(1)}\propto A^{\dagger}\psi_{0}^{(2)}$ computed using the converged
$\rho_{0}^{(2)}$ density.  As anticipated, it shows the proper symmetry and nodal position.

\begin{figure}[t]
{\includegraphics[width=0.5\columnwidth]{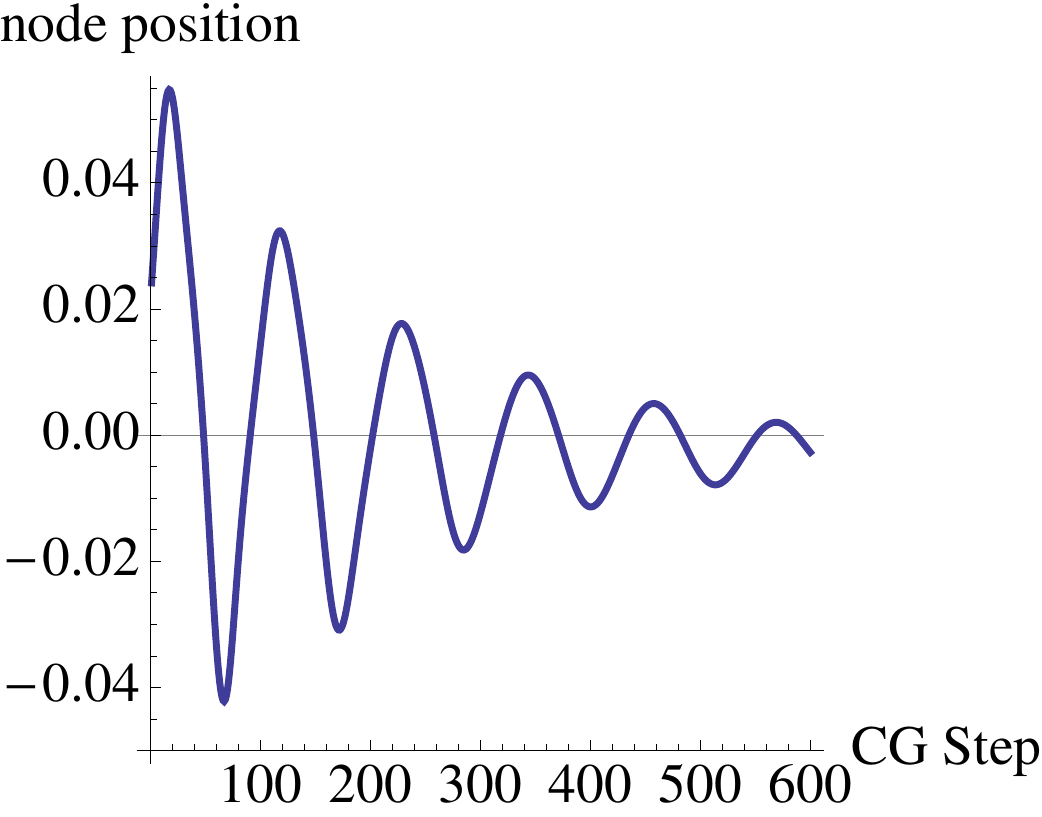}}
\caption{
Location of excited state node for the last 600 CG steps.
(From Ref.\cite{Bittner:2009jt})
 }\label{fig3}
\end{figure}

By symmetry, one expects the node to lie precisely at the origin.  However, since we have not imposed
any symmetry restriction or bias on our numerical method, the position of the node provides a sensitive test of the 
convergence of the trial density for $\rho_{0}^{(2)}$. In the example shown in Fig.\ref{fig3}, the location of the node oscillates
about the origin and appears to converge exponentially with number of CG steps. 
This is remarkably good considering that this is ultimately determined by the 
quality of the 3rd and 4th derivatives of $\rho_{o}^{(1)}$ that appear when computing the conjugate gradient of
$V_{2}$.  We have tested this approach on a number of other one-dimensional bound-state problems with similar success.

 \section{Extension of SUSY to multiple dimensions.}

 While SUSY-QM has also been explored for one dimensional,
non-relativistic quantum mechanical
problems\cite{baer:075024,Eides:1984rz,Andrianov:1985ty,Cooper:1995jt,Andrianov:1984nr,PhysRevA.47.2720},
thus far these studies have focused on the formal aspects and on
obtaining exact, analytical solutions for the ground state for
specific classes of problems. In several recent papers\cite{Bittner:2009jt,Kouri:2009hb,Kouri09,Kouri:2009xr},
we have begun
exploring the SUSY-QM approach as the basis of a general
computational scheme for bound state problems. Our initial
studies have been restricted to one dimensional systems (for which
there are, obviously, many powerful computational methods). In our
first paper, we found that SUSY-QM (combined with a new periodic
version of the Heisenberg-Weyl algebra) yields a robust, natural way
to treat an infinite family of hindered rotors.\cite{Kouri:2009xr}
Next we showed that the SUSY-QM leads to a general treatment of an
infinite family of anharmonic oscillators, such that highly accurate
excited state energies and wave functions could be obtained
variationally using significantly smaller basis sets than a
traditional variational approach requires\cite{Kouri:2009hb}.  Most
recently , we have considered a 1-D double well potential in which we
solved for the ground state energy and wave function using a VQMC
approach. Then using SUSY-QM, we (numerically) generated an auxiliary
Hamiltonian whose {\em nodeless} ground state is iso-spectral
(degenerate) with the first excited state of the original system
Hamiltonian. This ground state was also easily determined by VQMC,
yielding excellent accuracy for the first excited state
energy. \cite{Bittner:2009jt} Even more significant, by using the
charge operators naturally generated in the SUSY-QM approach, we also
obtained excellent accuracy for the first excited state wave function.
Furthermore, at no point did impose a fixed node or symmetry on the
excited state wave function and our calculation only involved working
with a nodeless ground state.

Of course, all this begs the question: Can this approach be
generalized to higher numbers of dimensions and to more than a single
particle?  There has been substantial effort in the past to do just
this.\cite{Eides:1984rz,Andrianov:1985ty,Cooper:1995jt,Andrianov:1984nr,leung:4802,rodrigues:125023,contreras-astorga:55,Andrianov:1987zl,Andrianov:1988yg,Andrianov:1984sf,A.A.Andrianov:1984qv,Andrianov:1986rm,0305-4470-35-6-305,Andrianov:2002gf,Das:1996sw,SIGMA}
However, to date, no such generalization has been found that is able
to generate all the excited states and energies even for so simple a
system as a pair of separable, 1-D harmonic oscillators (HO) or
equivalently, for a separable 2-D single HO. In our most recent,
unpublished work \cite{Kouri09}, we have succeeded in obtaining such a
generalization and showed that it does, in fact, yield the correct
analytical results for separable and non-separable problems.   In the next section, we 
present a succinct  summary of our approach. The major
question now is whether this formalism provides a basis for a robust,
computational method for determining excited state energies and wave
functions for large, strongly correlated systems using either QMC or
variational algorithms applied solely to nodeless ground state
problems.

\subsection{Difficulties in extending beyond one dimension}
To move beyond one dimensional SUSY, 
Ioffe and coworkers have explored the use of higher-order charge operators
\cite{A.A.Andrianov:1993bv,Andrianov:1995ve,0305-4470-35-6-305,Andrianov:2002gf},
and  Kravchenko  has explored the use of
 Clifford algebras\cite{Kravchenko}. 
 Unfortunately, this is difficult to do in general.  The reason being that
the Riccati factorization of the one-dimensional Schr\"odinger equation does not extend easily to 
higher dimensions.  
One remedy is write the charge operators as vectors $\vec{A} = (+\vec\partial + \vec W)$ 
and with  $\vec{A}^{+} = (-\vec\partial + \vec{W})^{\dagger}$ as the adjoint charge operator.  
The original Schr\"odinger operator is then constructed
as an inner-product
\begin{eqnarray}
H_{1} = \vec{A}^{+} \cdot \vec{A} .
\end{eqnarray}
Working through the vector product produces the Schr\"odinger equation
\begin{eqnarray}
H_{1}\phi = (-\nabla^{2} + W^{2}  - (\vec{\nabla}\cdot \vec{W})) \phi  = 0
\label{2dfactr}
\end{eqnarray}
and a Riccati equation of the form
\begin{eqnarray}
U(x) = W^{2} - \vec\nabla\cdot \vec{W}.
\end{eqnarray}
For a 2d harmonic oscillator, we would obtain a vector superpotential of the form
\begin{eqnarray}
\vec{W} =-\frac{1}{\psi_{0}^{(1)}} \vec\nabla\psi_{0}^{(1)} =  \left(x, y\right)  = (W_{x},W_{y})
\end{eqnarray}
Let us look more closely at the $\vec\nabla\cdot \vec{W}$ part.
If we use the form that $\vec{W} = - \vec{\nabla}\ln\psi$, then
$-\vec{\nabla}\cdot\vec{\nabla}\ln\psi = -\nabla^{2}\ln\psi$ which for the 
2D oscillator results in $\vec\nabla\cdot \vec{W} = 2$.
Thus, 
\begin{eqnarray}
W^{2} - \vec\nabla\cdot \vec{W } = (x^{2} + y^{2}) - 2 
\end{eqnarray}
which agrees with the original symmetric harmonic potential.
Now, we  write the scaled partner potential as
\begin{eqnarray}
U_{2}=W^{2} +  \vec\nabla\cdot \vec{W } =(x^{2} + y^{2})  + 2.\label{u2}
\end{eqnarray}
This is equivalent to the original potential shifted  by a constant amount.
\begin{eqnarray}
U_{2} = U_{1} + 4.
\end{eqnarray}
The ground state in this potential would be have the same energy as the
states  of the original potential with quantum numbers $n + m = 2$.    Consequently, 
even with the this na\"ive factorization, one can in principle obtain excitation energies for 
higher dimensional systems, but there is no assurance that one can reproduce the 
entire spectrum of states. 

The problem lies in the fact that  neither Hamiltonian $H_{2}$ nor its associated potential $U_{2}$ is given correctly by 
the form implied by Eq.~\ref{2dfactr} and Eq. \ref{u2}.   
Rather, the correct approach is to write the $H_{2}$
Hamiltonian as  a {\em tensor}  by taking the outer product of the charges
$\overline{H}_{2} = \vec{A} \vec{A}^{+}$ rather than as a scalar $\vec{A} \cdot\vec{A}^{+}$. 
At first this seems unwieldy and unlikely to lead anywhere since the wave function solutions  of
\begin{eqnarray}
\overline{H}_{2} \vec{\psi} = E \vec{\psi}
\end{eqnarray}
 are now
vectors rather than scalers.    However, rather than adding an undue complexity to the problem, it 
actually simplifies matters considerably. 
As we  demonstrate in a forthcoming paper, this tensor factorization preserves the SUSY algebraic 
structure and produces excitation energies for any $n-$dimensional SUSY system.  Moreover, this 
produces a scalar $\mapsto$ tensor $\mapsto$ scalar hierarchy as one moves to higher excitations.\cite{Kouri09}

\subsection{Vector SUSY}

We now give a brief summary of our new generalization of SUSY-QM to
treat higher dimensionality and more than one particle. Previous
attempts generally involved introducing additional,
``spin-like'' degrees of freedom.\cite{Andrianov:1985ty,Andrianov:1984nr,Andrianov:1987zl,Andrianov:1988yg,A.A.Andrianov:1984qv,Andrianov:1984sf,0305-4470-35-6-305,Andrianov:2002gf,A.A.Andrianov:1993bv,Dzhioev:2007,Andrianov:1995ve} 
In our approach, we make use of a
vectorial technique that can deal simultaneously with either higher
dimensions or more than one particle. In fact, the two problems are
dealt with in exactly the same manner. Therefore, for simplicity, we
consider a general $n$-dimensional distinguishable particle system
with orthogonal coordinates $\{x_{\mu}\}$.  The Hamiltonian is given
by \footnote{Our units are such that $\hbar^{2}/2m = 1$.}
 \begin{eqnarray}
H = -\nabla^{2 } + V_{0}(x_{1},\cdots x_{n})  
\end{eqnarray}
 and the nodeless ground state satisfies the Schr\"odinger equation,
\begin{eqnarray}
H \psi_{0}^{(1)} = E^{(1)}_{0}\psi_{0}^{(1)}. \label{eq2}
\end{eqnarray}
We now define a  ``vector super-potential'',  $\vec{W}_{1}$, with components
\begin{eqnarray}
W_{1\mu}  = - \frac{\partial}{\partial x_{\mu}} \ln\psi_0^{(1)}.
\end{eqnarray}
Then it is easily seen that the original Hamiltonian can be recast as 
\begin{eqnarray}
H_{1} = (-\nabla + \vec{W}_{1})\cdot (\nabla + \vec{W}_{1}) =
\vec{Q}_{1}^{+}\cdot \vec{Q}_{1}\label{eq5}
\end{eqnarray}
where the $\vec{Q}_{1}$ and $\vec{Q}_{1}^{+}$ are multi-dimensional
generalizations of the SUSY charge operators from
Eq. ~\ref{susy-hamiltonian}.  This defines our ``sector-1'' (or
``boson'') Hamiltonian and Eq.(\ref{eq2}) can be written as
\footnote{ In analogy with the original descriptions of SUSY, we refer
  to the partner pairs as ``boson'' and ``fermion'' sectors or less
  poetically as ``sector-1'', ``sector 2'', and so forth.}
 \begin{eqnarray}
H_{1}  \psi_{0}^{(1)} = E^{(1)}_{0}\psi_{0}^{(1)}
\end{eqnarray}  
One can show that the vector superpotential is related to the original
(scalar) potential via:
\begin{eqnarray}
V_{0} = \vec{W}_{1}\cdot \vec{W}_{1} - \nabla\cdot\vec{W}_{1}.
\end{eqnarray}
 The various components of the charge operators, $\vec{A}_{1}$ and
 $\vec{A}_{1}^{+}$ are defined by
\begin{eqnarray}
A_{1\mu} = \frac{\partial}{\partial x_{\mu}} + W_{1\mu} \,\,\&\,\,\,
A_{1\mu}^{+} = -\frac{\partial}{\partial x_{\mu}} + W_{1\mu}.
\end{eqnarray}
 Note that since these are associated with orthogonal degrees of
 freedom, the charge operators can be applied either by individual
 components or in vector form.

Next, consider the Schr\"odinger equation for the first excited state
of $H$.  We can write this using the charge operators as
\begin{eqnarray}
H_{1}\psi_{1}^{(1)} = E_{1}^{(1)}\psi_{1}^{(1)} = (\vec{A}_{1}^{+}\cdot\vec{A}_{1} + E_{0}^{(1)})\psi_{1}^{(1) }
\end{eqnarray}
 We apply  $\vec{A}_{1}$ to Equation (9):
\begin{eqnarray}
(\vec{A}_{1}\vec{A}_{1}^{+})\cdot \vec{A}_{1}\psi_{1}^{(1)} = ( E_{1}^{(1)} -E_{1}^{(0)} ) \vec{A}_{1}\psi_{1}^{(1)}
\end{eqnarray}
Here we identify $(\vec{A}_{1}\vec{A}_{1}^{\dagger})$ as a new,
auxiliary Hamiltonian.  It is important to note that this is
constructed from the outer or tensor product of the charge operators
rather than from inner or dot product as used in constructing $H_{1}$.
Its eigenvector, $ \vec{A}_{1}\psi_{1}^{(1)}$, is isospectral with the
excited state, $\psi_{1}^{(1)}$ of $H_{1}$ ( since $E_{0}^{(1)}$ is
known, determining $ ( E_{1}^{(1)} -E_{1}^{(0)} ) $ yields $
E_{1}^{(1)} $ ).  We therefore define the {\bf tensor} Hamiltonian for
the second sector as
\begin{eqnarray}
\tens{H}_{2} = \vec{A}_{1}\vec{A}_{1}^{\dagger} \label{eq11}
\end{eqnarray}
and {\bf vector} state function as
 \begin{eqnarray}
\vec\psi_{0}^{(2)} = \frac{1}{( E_{1}^{(1)} -E_{1}^{(0)} )
}\vec{A}_{1}\psi_{1}^{(1)}.
\end{eqnarray}
It is easy to show that the ground state energy of $\tens{H}_{2}$ is
related to the first excitation energy of the original Hamiltonianm
\begin{eqnarray}
E_{0}^{(2)} =  E_{1}^{(1)} -E_{1}^{(0)}. \label{eq13}
\end{eqnarray}
Furthermore, the ground state of $\tens{H}_{2}$ is also {\bf
  nodeless}. This has been explicitly shown to be true for the
separable 2-particle HOs considered earlier\cite{Kouri09}. Therefore, we
propose to apply both the VQMC and the standard variational methods to
determine $E_{0}^{(2)} $ and $ \vec\psi_{0}^{(2)}$. Of course, knowing
the second sector ground state energy also gives us the first excited
state energy of the original Hamiltonian
(Eq. (\ref{eq13})). Furthermore, we form the scalar product of
\begin{eqnarray}
\tens{H}_{2}\cdot \vec{\psi}_{0}^{(2)} = E_{0}^{(2)}
\vec{\psi}_{0}^{(2)} \label{eq14}
\end{eqnarray} 
with  $\vec{A}_{1}^{+}$ obtaining
\begin{eqnarray}
(\vec{A}_{1}^{+}\cdot \vec{A}_{1}) \vec{A}_{1}^{+}
  \vec{\psi}_{0}^{(2)} = E_{0}^{(2)} \vec{A}_{1}^{+} \cdot
  \vec{\psi}_{0}^{(2)}
\end{eqnarray}
Clearly, this is exactly
\begin{eqnarray}
H_{1} ( \vec{A}_{1}^{+} \cdot \vec{\psi}_{0}^{(2)} ) = E_{1}^{(1)} (
\vec{A}_{1}^{+} \cdot \vec{\psi}_{0}^{(2)} )
\end{eqnarray}
so we can conclude that
\begin{eqnarray}
\psi_{1}^{(1)} = \frac{1}{\sqrt{E_{0}^{(2)}}} ( \vec{A}_{1}^{+} \cdot
\vec{\psi}_{0}^{(2)} )
\end{eqnarray}

Thus we also obtain the excited state wave function without any
significant additional computational effort. This is because applying
the charge operator is much simpler than solving an eigenvalue problem
(it is a strictly linear operation). Evidence from our 1-D studies
indicates that the accuracy of the excited states obtained using the
SUSY-QM charge operator is significantly higher, for a given basis
set, than what is obtained variationally (or with QMC) from the
original Hamiltonian\cite{Kouri09,Kouri:2009hb}. This procedure can be continued as
follows. We define a sector-2 vector super-potential with components
\begin{eqnarray}
W_{2\mu} = \frac{\partial}{\partial x_{\mu}} \ln\psi_{0\mu}^{(2)} 
\end{eqnarray}
 Then it follows that
\begin{eqnarray}
\vec{A}_{2}\cdot \vec{\psi}_{0}^{(2)} = (\nabla + \vec{W}_{2})\cdot
\psi_{0}^{(2)} = 0
\end{eqnarray}
 so we can write
\begin{eqnarray}
\tens{H}_{2} = \vec{A}_{2}^{+} \vec{A}_{2} + E_{0}^{(2)} {\bf I}
\end{eqnarray}
 and Eq. (\ref{eq14}) is still satisfied. We form the scalar product
 of $ \vec{A}_{2}$ with the first excited state Schr\"odinger equation
 to obtain
\begin{eqnarray}
( \vec{A}_{2} \cdot \vec{A}_{2}^{+}) \vec{A}_{2} \cdot
  \vec{\psi}_{1}^{(2)} = E_{1}^{(2)} \vec{A}_{2}\cdot
  \vec{\psi}_{1}^{(2)}
\end{eqnarray} 
Then we define the sector 3 {\em scalar} Hamiltonian by
\begin{eqnarray}
H_{3} =  \vec{A}_{2} \cdot  \vec{A}_{2}^{+} + E_{0}^{(2)}
\end{eqnarray}
 with the ground state wave equation
\begin{eqnarray}
H_{3}\psi_{0}^{(3)} = E_{0}^{(3)} \psi_{0}^{(3)}.
\end{eqnarray}
It is easily seen that $E_{0}^{(3)} = E_{1}^{(2)} - E_{0}^{{2}}$.
This procedure continues until all bound states of the original
Hamiltonian are exhausted. It should also be clear that the sector 2
excited state wave function is obtained from the nodeless sector 3
ground state by applying $\vec{A}_{2}^{+}$ to it. Then the second
excited state for sector 1 results from taking the scalar product of $
\vec{A}_{1}^{+}$ with $\vec{\psi}_{1}^{(2)}$ .  The approach thus
leads to an alternating sequence of scalar and tensor Hamiltonians,
{\em but in all cases we need only determine nodeless ground states}.

There are two additional aspects of the tensor sector problem that
require discussion.  First we consider the validity of the
Rayleigh-Ritz variational principle. It is easily seen from
Eq. (\ref{eq11}) that $\tens{H}_{2}$ is a Hermitian operator.
Therefore, its eigenspectrum is real and its eigenvectors are
complete. With these facts in hand, the proof of the variational
principle follows the standard one in every detail. This is also true
for the Hylleraas-Undheim theorem.

Second, the QMC method is also directly applicable to the tensor
sector problem. For the example discussed above, we note that the
energy is given by
 \begin{eqnarray}
 E_{trial} = \frac{\int d\tau \vec\psi_{trial}\cdot \tens{H}_{2} \cdot
   \vec{\psi}_{trial}}{\int d\tau \vec\psi_{trial}\cdot
   \vec{\psi}_{trial}}
 \end{eqnarray}
We next note that the integral can be expanded in terms of its
components as
\begin{eqnarray}
E_{trial} = \frac{\sum_{\mu\nu}\int d\tau (\psi_{\mu,trial}
  H_{2,\mu\nu} \psi_{\nu,trial} )} {\sum_{\mu}\int d\tau(
  \psi_{\mu,trial})^{2}}.
\end{eqnarray}
It is then clear that each separate integral can be evaluated by
QMC. For example, the $\mu\ne \nu$ cross term is divided and
multiplied by
\begin{eqnarray}
\psi_{\mu,trial}\int d\tau \psi_{\nu,trial}\psi_{\mu,trial}.
\end{eqnarray}
Then the sampling is done relative to the mixed probability
distribution,
\begin{eqnarray}
P_{\mu\nu} = \frac{ \psi_{\mu,trial}\psi_{\nu,trial}}{\int d\tau  \psi_{\mu,trial}\psi_{\nu,trial} }.
\end{eqnarray}
A similar expression applies to each term in the energy expression and
the evaluation would need to be performed self-consistantly.

Thus far, we have developed a formalism that appears to be suitable
for extending the SUSY-QM technique to higher dimensional systems.  We
believe the approach we have outlined above will provide the
mathematical basis for a number of potentially interesting theoretical
results.  Moreover, we anticipate that when combined with either
variational or Monte Carlo methods, our multi-dimensional extension of
SUSY-QM will facilitate the calculation of accurate excitation
energies and excited state wave functions.

 \section{Outlook}

I presented a number of avenues we are actively pursuing with the
goal of using SUSY-QM or SUSY-inspired-QM to solve problems that are difficult to solve using more 
conventional approaches.  In addition to what I have discussed here we exploring the use of the Riccati equation to 
solve quantum scattering problems. 
It is as if one of the co-authors of this paper (DJK) has 
come full-circle since one of his first papers 
concerned  solving the 
Hamilton-Jacobi for the action integral in quantum scattering,\cite{kouri:1919}\footnote{Coincidentally, Ref. \cite{kouri:1919}  appeared in the J. Chem. Phys. issue immediately before the birthday of the other author of this paper.
There appears to be some interesting Karma at work here.} 
$$
iS/\hbar = -\int_{r_{o}}^{r}W(r')dr'.
$$
The integrand in this last equation is the SUSY super-potential. Furthermore, there is a connection between 
our work and the complex-valued quantum trajectories studied by Wyatt and Tannor and their respective co-workers.

\begin{acknowledgements}

This work was supported in part by the National Science Foundation 
(ERB: CHE-0712981) and the Robert A. Welch foundation (ERB: E-1337, DJK: E-0608).
The authors also acknowledge Prof. M. Ioffe for comments regarding the 
extension to higher dimensions. 
\end{acknowledgements}

\end{document}